\documentstyle[12pt]{article}
\newcommand{\md}{\mbox{d}}
\begin{document}\hbadness=10000
\thispagestyle{empty}
\pagestyle{myheadings}\markboth{H.-Th. Elze}
{The Functional Derivation of Master Equations}
\title{The Functional Derivation of Master Equations\footnote
{A preliminary version will be published in Proceedings of Hadron Physics VI, E. Ferreira et al., eds. (World Scientific, Singapore, 1999).} 
}
\author{$\ $\\
{\bf Hans-Thomas Elze} \\ \ \\
Universidade Federal do Rio de Janeiro, Instituto de
F\'{\i}sica \\
\,\,\,\,\,\,Caixa Postal 68.528, 21945-970 Rio de Janeiro, RJ, Brazil 
}
\date{May 1999} 
\maketitle 
  
\begin{abstract}
{\noindent     
Master equations describe the quantum dynamics of 
open systems interacting with an environment. They play an 
increasingly important role in understanding the 
emergence of semiclassical behavior and the generation 
of entropy, both being related to quantum decoherence. 
Presently we derive the exact master equation 
for a homogeneous scalar Higgs or inflaton like field coupled to an environment field. It is represented here by an infinite set of harmonic oscillators with a given spectral density.  
Our nonperturbative result follows directly from the path integral representation of the density matrix propagator. Applications and generalizations are discussed.        
}   
\end{abstract}

\section{Introduction}
Often the physical system under consideration cannot be 
studied in isolation. In such cases the interaction with the 
environment has to be taken into account from the outset, 
despite the fact that one 
would like to eliminate these `uninteresting' degrees of freedom 
from the dynamical description of the proper open subsystem one is 
really interested in. Prominent classical examples are  
the motion of a massive body under the influence of friction or 
the Brownian motion, where the idealization of an isolated 
system clearly misses the relevant effects. It is not hard 
to imagine corresponding quantum mechanical situations, most 
notably the effect of Ohmic resistance to the electron current 
in a wire, and similar other dissipative systems with an interplay 
of transport and relaxation phenomena. The results of research 
in this field of the so-called quantum Brownian motion have been 
reviewed, e.g., in 
Ref.\,\cite{Grabert88}, where numerous further references can 
be found. 
  
Recently open systems have been studied 
also in the context of quantum field theories, 
with applications ranging from high-energy physics to cosmology. 
The underlying theme herein is the study of quantum decoherence, 
i.e. the emergence of semiclassical properties in complex 
quantum mechanical systems. In terms of the density matrix of 
the open subsystem one would like to understand how a more or less pure initial state evolves into an 
impure mixture. Clearly this question is intimately related 
to the problem of entropy generation and possibly thermalization 
in quantum systems. The references  
illustrate the different physical aspects of the environment 
induced decoherence \cite{UnZu89,BaCald91,I95,LoMa96}, which also may solve the (in)famous 
problem of the `wave function collapse' \cite{Zu91,Teg93}. 

In any case, let us recall the general structure of all master 
equations as expressed in terms of density matrices. -- The overall   
density matrix $\hat\varrho$ describing the subsystem plus its environment 
can be expanded in a complete set of states 
${\Psi_n}$\,, which obey the time dependent Schr\"odinger equation, 
$\hat\varrho =\sum_nc_n|\Psi_n\rangle\langle\Psi_n|$\,. Then, employing 
the Schr\"odinger equation, the equation of motion for the 
density matrix follows immediately: 
$i\partial_t\hat\varrho =[\hat H_{SE},\hat\varrho ]$\,, 
where $\hat H_{SE}$ denotes the total system plus environment 
Hamiltonian; our units are such that $\hbar =c=k_B=1$\,. The aim of all derivations of master equations is 
to reduce the equation for $\hat\varrho$ to an equation for 
the subsystem density matrix $\hat\rho$\,, which is obtained by eliminating the environment degrees of freedom, $\hat\rho\equiv\mbox{Tr}_{E}\hat\varrho$\,. The result, generally, 
is of the form  
\begin{equation} \label{master} 
i\partial_t\hat\rho =[\hat H,\hat\rho ]+F[\hat\rho ]
\;\;, \end{equation} 
where $\hat H$ is the subsystem Hamiltonian and the remaining term 
describes the dissipative and diffusive effects of the integrated out environment; as we shall see in the following, it may have a complicated 
time dependent (memory) structure possibly involving derivatives of 
the subsystem density matrix $\hat\rho$\,.  

In order to proceed, we consider a simple model motivated by studies 
of the inflationary universe \cite{GuPi85,CoMaSq89}. The semiclassical homogeneous field equation for the scalar `inflaton' $\Phi$ 
is 
\begin{equation} \label{inflaton} 
\ddot\Phi +3H\dot\Phi +\Gamma\dot\Phi =-\frac{\partial V}{\partial\Phi}
\;\;, \end{equation} 
where $H\equiv \dot R/R$ is the Hubble `constant' and the corresponding 
friction term here describes the damping due to the expansion of the 
universe; similarly, the term involving the decay rate $\Gamma$ 
describes the essentially dissipative coupling to other particles; 
the r.h.s. contains the ordinary force term due to the selfinteraction,  
e.g. $V(\Phi )\equiv -\frac{1}{2}\mu^2\Phi^2+\frac{1}{4}\lambda\Phi^4$\,.  

Whereas the fate of the inflaton field driving the early universe 
expansion has been much studied including 
its quantum as well as thermal fluctuations, we would like to draw 
attention to another aspect of Eq.\,(\ref{inflaton}). It could  
arise from a typical semiclassical saddle-point approximation to 
a path integral describing the evolution of $\Phi$ in a higher-dimensional space-time environment \cite{CoMaSq89,ACF87}. In this case, the interesting 
long-time behavior of the inflaton field will be modified by the 
influence of quantum fluctuations in the higher-dimensional background 
space. 

In the following, leaving the wider scope for now, we consider as our starting point 
the classical action 
\begin{equation} \label{action} 
S_{SE}=\int\md t\left\{\frac{1}{2}\dot\Phi^2-V(\Phi) 
+\sum_{n=1}^N\frac{m_n}{2}[\dot\phi_n^{\;2}
-\omega_n^{\;2}(\phi_n-\Phi )^2]\right\}
\;\;, \end{equation}
i.e. coupling the homogeneous scalar field to an 
environment of $N\rightarrow\infty$ oscillators,  
which is equivalent 
to a single radiation field \cite{UnZu89,BaCald91}. 
This Feynman-Vernon or Caldeira-Leggett type of model 
represents a large variety of different physical 
situations depending on the distribution of oscillator 
masses (dimensionless) and frequencies, and the potential, 
$m_n,\,\omega_n$, and $V(\Phi )$, respectively. 

Similar models have   
been studied in numerous special cases before, see 
e.g. Refs.\,\cite{Grabert88,BaCald91,I95}. In particular, 
the effective $\Phi$-equation obtained in the 
saddle-point approximation after integrating out 
the environment degrees of freedom 
exactly (cf. Sec.\,2) assumes the form 
of Eq.\,(\ref{inflaton}) for the so-called Ohmic environment 
\cite{Grabert88}, cf. Sec.\,3.
  
In distinction to most earlier work, we do not attempt  
to solve the quantum mechanical time evolution problem for the  
density submatrix of the $\Phi$-field in a special case. 
Rather our aim is to derive its master equation employing 
the path integral representation; see Ref.\,\cite{UnZu89} for comparison, where a master equation was derived  
using the canonical quantization of a somewhat different model. We 
believe that the nonperturbative method presented here lends itself to 
generalizations for more genuine field theoretic problems,
i.e. involving propagating fields \cite{LoMa96}, provided the 
relevant influence 
functional (cf. Sec.\,2) can be calculated.      
     
\section{From the Influence Functional to the Master Equation}
To begin with, since the above action, 
Eq.\,(\ref{action}), is quadratic in the oscillator coordinates 
and velocities $\phi_n,\,\dot\phi_n$\,, they can be eliminated exactly 
from the path integral(s) describing the time evolution of the complete 
system \cite{Grabert88}. In the following we neglect initial state correlations between the proper subsystem and the 
environment degrees of freedom for simplicity; their  
incorporation does not add a new element to our derivation.

It turns out that the result of the Gaussian integrations can be 
specified completely in 
terms of the thermal noise and the dissipation kernels, 
$\nu$ and $\eta$, respectively,
\begin{equation} \label{kernels}
\nu (s)\equiv\int _0^\infty\md\omega\; I(\omega )\;
\coth (\frac{\beta\omega}{2})
\cos (\omega s)\;\; ,\;\;\;\eta (s)\equiv -\int _0^\infty\md\omega
\; I(\omega )\;\sin (\omega s)\equiv\frac{\md}{\md s}\bar\eta (s)
\;\;, \end{equation} 
where $\beta\equiv T^{-1}$ denotes the inverse temperature.  
The spectral density here is defined by  
\begin{equation} \label{spectrald} 
I(\omega )\equiv\frac{1}{2}\sum _{n=1}^Nm_n\omega _n^3
\;\delta (\omega -\omega _n)
\;\;. \end{equation} 
It characterizes the particular environment.
Assuming a quasi-continuous
distribution of environmental oscillators, spectral densities
$I(\omega )\propto\omega ^k$ (for sufficiently small $\omega$)
with $k>0$ have often 
been considered before, cf. \cite{Grabert88,UnZu89,I95}, and  
have been  
called sub-Ohmic ($k<1$), Ohmic ($k=1$), and supra-Ohmic ($k>1$), respectively. The electron with its radiation field, for example,
amounts to a particle in a supra-Ohmic environment with $k=3$ \cite{BaCald91}.
 
In this way, the following representation for the $\Phi$-`field' 
density matrix at time $t$ in terms of the initial one ($t=0$) is obtained: 
\begin{equation} \label{densityt} 
\rho (\Phi,\Phi',t)=\int\md\Phi_i\md\Phi_i'\;
J(\Phi,\Phi ',t;\Phi_i,\Phi_i',0)\;
\rho (\Phi_i,\Phi_i',0)
\;\;, \end{equation} 
with the propagator determined by two remaining path integrals \cite{Grabert88},
\begin{equation} \label{propagator}
J(\Phi,\Phi ',t;\Phi_i,\Phi_i',0) 
=\frac{1}{Z}\int_{\Phi_i,\Phi_i',0}^{\Phi,\Phi ',t} 
{\cal D}q{\cal D}q'\;
\exp\{i(S[q]-S[q'])\}
\exp\{-{\cal F}[q,q',t]\}
\;\;, \end{equation}
and with the indicated boundary conditions, $q(0)\equiv\Phi_i$, $q(t)\equiv\Phi$, and
for $q'$ analogously. Here $Z$ is a (generally time-dependent) 
normalization factor, such that the 
total probability to find the subsystem  
is preserved (see the first of Refs.\,\cite{I95} for a detailed discussion of general properties of open quantum systems, the subsystem density 
matrix in particular, as well as an explicit calculation); we omit $Z$ 
in the following, since it can always be restored by properly normalizing 
$\hat\rho$ in the very end. Furthermore, the action  
$S$ is the part of $S_{SE}$, Eq.\,(\ref{action}),   
which only depends on the field $\Phi$\,, and      
the Feynman-Vernon influence functional ${\cal F}$ incorporates the
effect of the integrated out environment on the subsystem. 
In the present case it is given by 
\begin{eqnarray} \label{influencef}
{\cal F}[q,q',t]&=&\int _0^t\md s\int _0^s\md u\;
[q(s)-q'(s)]\left\{\nu (s-u)[q(u)-q'(u)]
+i\eta (s-u)[q(u)+q'(u)]\right\} \nonumber \\
&\;&+i\bar\eta (0)\int _0^t\md s\; [q^2(s)-q'^2(s)]
\;\;. \end{eqnarray} 
Clearly, the spectral distribution $I(\omega )$ has to vanish sufficiently
fast in the infrared, in order to avoid a
divergence of $\bar\eta (0)$, cf. Eqs.\,(\ref{kernels}).  
We observe that the last two terms in Eq.\,(\ref{influencef}) simply renormalize the quadratic 
parts of the potentials in $S[q]-S[q']$ in Eq.\,(\ref{propagator}); therefore, we omit these terms in 
the following and add them again in the end, see Eqs.\,(\ref{potentialr}).   

Having set the stage with Eqs.\,(\ref{densityt})--(\ref{influencef}), 
we derive the master equation by suitably evaluating 
the time derivative $\partial_t\rho =
\lim_{\epsilon\rightarrow 0}[\rho (t+\epsilon )-
\rho (t)]/\epsilon$\,. We calculate to $O(\epsilon )$ the 
r.h.s. of $\rho (\Phi,\Phi ',t)+\dot\rho (\Phi,\Phi ',t)\epsilon\simeq\rho (\Phi,\Phi ',t+
\epsilon )$ by going through the following steps: 

\noindent {\bf i.}  
We split the path integrals going from $0$ to $t+\epsilon$ at time $t$, cf. Eqs.\,(\ref{densityt})--(\ref{influencef}), by inserting the intermediate 
`points' $z,z'$ and integrating -- corresponding to 
inserting a complete set of states in the formula 
$\exp\{ i\hat H(t+\epsilon )\}=\int\md z\;\exp\{ i\hat H\epsilon\}|z\rangle\langle z|\exp\{ i\hat Ht\}$\,.

\noindent {\bf ii.}
We parametrize the (infinitesimal) propagation 
from $t$ to $t+\epsilon$ by straight line paths, 
\begin{equation} \label{paths} 
\tilde q(s)\equiv z+\frac{\Phi-z}{\epsilon}
(s-t)\;\;,\;\;\;
\dot{{\tilde q}}(s)=\frac{\Phi-z}{\epsilon}\;\;,\;\;\; t\leq s\leq t+\epsilon
\;\;, \end{equation}
and analogously $\tilde q'$\,; this parametrization 
becomes exact as $\epsilon\rightarrow 0$\,.

\noindent {\bf iii.}
We introduce $\chi\equiv z-\Phi$\,, and $\chi '$ 
correspondingly, and replace the 
integrations over $\md z,\,\md z'$ by the ones over $\md\chi ,\,\md\chi '$\,. -- The action for the propagation from $t$ to $t+\epsilon$ becomes 
\begin{equation} \label{infiniaction}
S\simeq\int_t^{t+\epsilon}
\md s\left\{\frac{1}{2}\dot{{\tilde q}}^2
-V(\tilde q)\right\}=\frac{1}{2}\frac{\chi^2}{\epsilon}
-\epsilon V(\Phi+\xi\chi )
\;\;, \end{equation}
where $\xi\in [0,1]$\,; $S[q']$ is calculated 
analogously. -- 
Similarly, we evaluate the influence functional, 
\begin{eqnarray} \label{influencef1}
&\;&{\cal F}[q,q',t+\epsilon ]={\cal F}[q,q',t]
+\int_t^{t+\epsilon }\md s\int_0^s\md u\;\dots 
\nonumber \\[2ex]
&\;&={\cal F}[q,q',t]
+\int_t^{t+\epsilon }\md s\int_0^t\md u\;\dots\;+
O(\epsilon^2) 
\nonumber \\[2ex] 
&\;&={\cal F}[q,q',t]+O(\epsilon^2,\epsilon\chi ,\epsilon\chi ')
\nonumber \\ 
&\;&\;\;\; +\epsilon (\Phi+\chi -\Phi '-\chi ')
\int_0^t\md u\;\{\nu (t-u)[q(u)-q'(u)]+i\eta (t-u)[q(u)+q'(u)]
\} 
\;\;, \end{eqnarray}
where we omitted the last two terms from Eq.\,(\ref{influencef}). 

\noindent {\bf iv.}
Expanding consistently to $O(\epsilon^2,\epsilon\chi ,
\epsilon\chi ')$, we obtain 
\begin{eqnarray} \label{densityteps} 
&\;&\rho (\Phi,\Phi ',t+
\epsilon )\simeq\int\md\chi\md\chi '\md\Phi_i\md\Phi_i'\;
\frac{1}{2\pi\epsilon}\exp\{\frac{i}{2\epsilon}(\chi^2
-\chi '^2)\}\left (1-i\epsilon V(\Phi)+i\epsilon V(\Phi ')\right ) 
\nonumber \\[1ex] 
&\;&\cdot\int_{\Phi_i,\Phi_i',0}^{\Phi +\chi ,\Phi '+\chi ',t} 
{\cal D}q{\cal D}q'\;
\exp\{i(S[q]-S[q'])\}
\exp\{-{\cal F}[q,q',t]\}\rho (\Phi_i,\Phi_i',0)
\nonumber \\ 
&\;&\cdot\left (1-\epsilon\{\Phi+\chi -\Phi '-\chi '\}
\int_0^t\md u\{\nu (t-u)[q(u)-q'(u)]+i\eta (t-u)[q(u)+q'(u)]
\}\right ) 
, \end{eqnarray}
where a possibly unexpected normalization factor $(2\pi\epsilon )^{-1}$ 
appears on the r.h.s.; analogously to the factor 
$(2\pi\epsilon )^{-1/2}$ in the derivation 
of the Schr\"odinger equation from the path integral, where one 
employs only one intermediate coordinate integration, it is  
necessary here, in order that Eq.\,(\ref{densityteps}) becomes an 
identity in the limit $\epsilon\rightarrow 0$\,. 

\noindent {\bf v.}
We observe that without the last factor under 
the remaining path integrals they would yield simply 
$\rho (\Phi+\chi ,\Phi_f'+\chi ',t)$\,. Furthermore, 
the Gaussian integrals over $\md\chi ,\,\md\chi '$ effectively require $\chi^2,\chi '^2\simeq\epsilon$\,. Therefore, properly expanding and collecting terms to 
$O(\epsilon )$, we obtain after performing these 
integrals the fully general result corresponding to  
the action of Eq.\,(\ref{action}): 
\begin{eqnarray} \label{densitygen} 
&\;&\partial_t\rho (\Phi ,\Phi ',t)= 
-\frac{1}{2i}[\partial_\Phi^{\;2} -\partial_{\Phi '}^{\;2}]
\rho (\Phi ,\Phi ',t)+\frac{1}{i}[V(\Phi )-V(\Phi ')]
\rho (\Phi ,\Phi ',t)
\nonumber \\ [1ex] 
&\;&\;\;-(\Phi -\Phi ')\int\md\Phi_i\md\Phi_i'
\int_{\Phi_i,\Phi_i',0}^{\Phi ,\Phi ',t} 
{\cal D}q{\cal D}q'\;
\exp\{i(S[q]-S[q'])\}
\exp\{-{\cal F}[q,q',t]\}\rho (\Phi_i,\Phi_i',0)
\nonumber \\ 
&\;&\;\;\cdot
\int_0^t\md u\;\{\nu (t-u)[q(u)-q'(u)]+i\eta (t-u)[q(u)+q'(u)]
\} 
\;\;, \end{eqnarray}
where, of course, we cancelled the leading terms between 
the l.h.s. and the r.h.s. and divided the remaining ones by $\epsilon$, before taking the limit $\epsilon\rightarrow 0$\,. We recall that the quadratic potential terms have 
to be renormalized,  
\begin{equation} \label{potentialr}
V(\Phi )\equiv 
-\frac{1}{2}\mu^2\Phi^2+\frac{1}{4}\lambda\Phi^4
\;\;\longrightarrow\;\;
V(\Phi )_r=-\frac{1}{2}
[\mu^2-2\bar\eta (0)]\Phi^2+\frac{1}{4}\lambda\Phi^4
\;\;, \end{equation} 
i.e., one has to perform a mass shift here. 
 
This completes our derivation of the master equation, which 
conforms with the general structure of Eq.\,(\ref{master}). 
Generally, we are still left with two  
formidable path integrals to be evaluated. However, 
depending on the actual form of the noise and dissipation 
kernels, the above result, Eq.\,(\ref{densitygen}) is a useful starting point for various approximation techniques, such as perturbation theory, 
mean field theory, variational techniques, or exact evaluations. 

The latter can be carried out  
in particular, if the action $S$ is at most quadratic in 
$\Phi ,\dot\Phi$ \cite{HuPazZh92}, or else, if the kernels have zero range in 
time corresponding to the absence of non-Markovian memory effects. In the following section 
we will show how the previously studied  
Ohmic environment can be treated with the help 
of the above results.      

\section{Example: Ohmic Environment}   
In order to illustrate the explicit form of the master equation 
(\ref{densitygen}), we choose the simplest example of an 
Ohmic environment \cite{Grabert88}. We begin with a quasi-continuous spectral 
density of the form \cite{HuPazZh92}
\begin{equation} \label{spectrald1} 
I(\omega ) =\frac{\gamma}{\pi}\omega
\left (\frac{\omega}{\tilde\omega}\right )^{k-1}
\exp\{-\omega^2/\Lambda^2\} 
\;\;, \end{equation} 
where $\Lambda$ is a high-frequency cut-off, $\tilde\omega$ 
another frequency scale, and $\gamma$ denotes the effective coupling constant between the $\Phi$-field and the 
environment oscillators, cf. Eqs.\,(\ref{action}),\,(\ref{spectrald}).    
 
Then, considering the high-temperature limit 
$T\gg\Lambda\ge\omega$ and especially the 
Ohmic environment with $k=1$\,, we obtain the 
thermal noise and 
dissipation kernels according to Eqs.\,(\ref{kernels}) 
\begin{equation} \label{kernels1} 
\nu (s)=2\gamma T\delta (s)\;\;,\;\;\;
\eta (s)=\gamma\delta '(s)
\;\;, \end{equation} 
where the prime denotes the derivative of the $\delta$-function 
in the last equation.   
Using these kernels in Eq.\,(\ref{densitygen}), the path 
integrals simply reproduce $\rho (\Phi ,\Phi ',t)$ and the 
resulting equation becomes essentially local in time, 
\begin{eqnarray} \label{densityOhm} 
&\;&i\partial_t\rho (\Phi ,\Phi ',t)=
\nonumber \\[1ex]
&\;&\left\{ 
-\frac{1}{2}[\partial_\Phi^{\;2} -\partial_{\Phi '}^{\;2}]
+V(\Phi )-V(\Phi ')
-i\gamma\{ 2T[\Phi -\Phi ']^2
+[\Phi -\Phi '][\partial_\Phi -\partial_{\Phi '}]\}
\right\}\rho (\Phi ,\Phi ',t)
\nonumber \\ 
&\;&+\gamma\delta (t)[\Phi -\Phi']\int\md\Phi_i\md\Phi_i'\;
[\Phi_i+\Phi_i']\rho (\Phi_i,\Phi_i',0)  
\;\;. \end{eqnarray}
We remark that herein a divergent term $-\gamma\delta (0)[\Phi^2-\Phi '^2]\rho (\Phi ,\Phi ',t)$ precisely 
cancelled the mass shift of Eq.\,(\ref{potentialr}); this 
has been the reason to introduce the counterterm 
$\sum_{n=1}^{N}m_n\omega_n^{\;2}\Phi^2/2$ into the action, 
Eq.\,(\ref{action}), in the first place. Furthermore, the 
singular initial value term in the last line of Eq.\,(\ref{densityOhm}) arises similarly from  
evaluating by partial integration the last term proportional to $\eta (t-u)=\gamma\delta '(t-u)$ in  Eq.\,(\ref{densitygen}). This contribution may vanish for 
particular initial conditions. However, it has not been noticed in earlier 
work on the Ohmic environment by Caldeira and Leggett and 
by Unruh and Zurek \cite{UnZu89} as summarized in Ref.\,\cite{HuPazZh92}. It produces an initial `jolt' 
which has been discussed   
in these studies of quantum decoherence. 
  
In order to cast the master equation (\ref{densityOhm}) 
into the form of a sometimes more useful Fokker-Planck 
type transport equation, we introduce the Wigner transform
\begin{equation} \label{Wigner} 
W(\Phi ,\Pi ,t)\equiv\int\md\phi\;\exp\{i\Pi\phi\}
\rho(\Phi +\textstyle{\frac{\phi}{2}},
\Phi -\textstyle{\frac{\phi}{2}},t) 
\;\;. \end{equation} 
With this definition and assuming that the potential 
$V(\Phi )$ is a fourth order polynomial, we obtain the 
transport equation 
\begin{eqnarray} \label{transport} 
\partial_tW(\Phi ,\Pi ,t)&=&
\left\{\Pi\partial_\Phi -V'(\Phi )\partial_\Pi
+\frac{1}{24}V'''(\Phi )\partial_\Pi^{\;3}
+4\gamma\{\partial_\Pi\Pi +T\partial_\Pi^2\}
\right\} W(\Phi ,\Pi ,t)
\nonumber \\ [1ex]
&\;&-4\pi\gamma\delta (t)\delta '(\Pi )\int\md\Phi'\;
\Phi' W(\Phi' ,0,0) 
\;\;, \end{eqnarray} 
where the derivatives act on everything to the right and 
the primes on $V(\Phi )$ denote differentiation w.r.t. $\Phi$\,. 

The various terms arising in Eq.\,(\ref{transport}) have the 
following significance: a) the l.h.s. together with 
the first two terms on the r.h.s. describe the classical 
Liouville flow of the phase space distribution $W(\Phi ,\Pi )$\,; b) the third term presents the quantum correction 
(for a non-polynomial potential one obtains here a well 
known sine series in powers of $\partial_\Phi\partial_\Pi$\,); c) the fourth and fifth 
term, respectively, incorporate the previously obtained 
dissipative and diffusive contributions due to the 
(integrated out) Ohmic environment; d) the highly singular 
last term indeed induces an initial `jolt' in the Wigner 
function, 
\begin{equation} \label{Wignerjolt} 
W(\Phi ,\Pi ,t=0_+)=W(\Phi ,\Pi ,t=0_-)
-4\pi\gamma\delta '(\Pi )\int\md\Phi'\;
\Phi' W(\Phi' ,0,t=0_-) 
\;\;. \end{equation}    
With this observation we conclude our (re-)derivation 
of more or less known features for the Ohmic environment 
and turn to a brief discussion of future applications 
of the method and main result presented in Sec.\,2.  

\section{Discussion} 
We derived the master equation for a homogeneous scalar
field interacting with an environment composed of 
harmonic oscillators, which may alternatively be 
viewed as a free radiation field \cite{BaCald91}, 
directly from the path integral representation of the 
density matrix propagator. In particular, we allowed 
for arbitrary self-interactions of the scalar field. 
We remark that our results apply equally 
well to a single nonrelativistic particle in an arbitrary 
external potential interacting with the environment, 
i.e. the case usually studied in quantum Brownian 
motion \cite{Grabert88}; then, the variable $\Phi$ is 
to be interpreted as the particle coordinate.    

As an immediate application of the results presently 
obtained we plan to study the phenomenon of semiquantum 
chaos in more realistic situations. 

Semiquantum chaos as 
described in Ref.\,\cite{BlEl96} (see also further 
references therein) arises in classically regular systems 
in the semiclassical regime. In particular, the time dependent 
Hartree-Fock approximation (or large-N expansion etc.) to the full quantum evolution, 
which in turn is regular, generally leads to nonlinearly 
coupled equations of motion for the one- and two-point functions. They have been shown to produce chaotic 
behavior under various circumstances. 

Our aim is to show that the interaction with a 
suitably prepared environment, e.g. involving 
solid state quantum dots or QED microwave cavities, can 
drive the system, e.g. conduction band electrons or Rydberg atoms, into the semiclassical limit where semiquantum 
chaos is expected to manifest itself.       

As is obvious from the 
general result, Eq.\,(\ref{densitygen}), or the 
more detailed calculation for the Ohmic environment 
which led to Eq.\,(\ref{densityOhm}) and the 
Fokker-Planck type transport equation (\ref{transport}), 
these master equations all are linear in the density 
matrix or Wigner function. An important generalization 
of the approach of Sec.\,2 would be to study the 
coupling between propagating fields. In this case we 
also expect nonlinear contributions to the effective `one-body' 
master equation which are due to scattering 
terms. We hope to come back to the problem of propagating 
fields in the 
near future with a study of the entropy production 
due to bremsstrahlung emission during the initial stopping
phase of high-energy heavy-ion collisions \cite{I95}. 
  
\subsection*{Acknowledgement} 
I thank C. E. Aguiar, T. Kodama and J. Rafelski for stimulating discussions 
and the CNPq (Brazil) for support, grant 
CNPq-300758/97-9.

\end{document}